\documentclass{emulateapj}
\newcommand{\bl}[1]{\mbox{\boldmath$ #1 $}}
\newcommand{\beq}{\begin{equation}}
\newcommand{\eeq}{\end{equation}}

\newcommand{\Alf}{Alfv\'en\ }
\newcommand{\Alfvc}{Alfv\'enic\ }
\newcommand{\alf}{Alfv\'en}
\newcommand{\cmc}{~{\rm cm}^{-3}}

\newcommand{\cs}{c_{\rm s}}
\newcommand{\pc}{\rm pc}
\newcommand{\csq}{c_{\rm s}^2}
\newcommand{\ceff}{C_{\rm eff}}
\newcommand{\ceffsq}{C_{\rm eff}^2}
\newcommand{\vai}{V_{\rm A,0}}
\newcommand{\vms}{V_{\rm MS,0}}
\newcommand{\vph}{V_{\rm MT,0}}
\newcommand{\vrms}{v_{\rm rms}}
\newcommand{\sign}{\sigma_{\rm n}}
\newcommand{\signi}{\sigma_{\rm n,0}}
\newcommand{\rhoni}{\rho_{\rm n,0}}
\newcommand{\Bref}{B_{\rm ref}}
\newcommand{\Beq}{B_{z,\rm eq}}
\newcommand{\tnii}{\tau_{\rm ni,0}}
\newcommand{\tniitil}{\tilde{\tau}_{\rm ni,0}}
\newcommand{\mui}{\mu_0}
\newcommand{\Pext}{P_{\rm ext}}
\newcommand{\Pexttil}{\tilde{P}_{\rm ext}}
\newcommand{\Ekin}{E_{\rm kin}}
\newcommand{\Ekini}{E_{\rm kin,0}}
\newcommand{\nn}{n_{\rm n}}

\slugcomment{Accepted by The Astrophysical Journal}
\shorttitle{Long-lived Modes}
\shortauthors{Shantanu Basu and Wolf B. Dapp}
\begin{document}

\title{Long-lived Magnetic-Tension-Driven Modes in a Molecular Cloud}

\author{Shantanu Basu and Wolf B. Dapp}
\affil{Department of Physics and Astronomy, University of Western Ontario,
London, Ontario, N6A 3K7, Canada}
\email{basu@uwo.ca}
\email{wdapp@uwo.ca}

\begin{abstract}{}
We calculate and analyze the longevity of magnetohydrodynamic (MHD) wave 
modes that occur in the plane of a magnetic thin sheet. Initial 
turbulent conditions applied to a magnetically subcritical cloud 
are shown to lead to
relatively rapid energy decay if ambipolar diffusion
is introduced at a level corresponding to partial ionization
primarily by cosmic rays. However, in the flux-freezing limit, as 
may be applicable to photoionized molecular cloud envelopes,
the turbulence persists at ``nonlinear'' levels in comparison
with the isothermal sound speed $\cs$, with one-dimensional rms 
material motions in the range of
$\approx 2\,\cs - 5\,\cs$ for cloud sizes in the range of 
$\approx 2\,\pc - 16\,\pc$. These fluctuations persist indefinitely, 
maintaining a significant portion of the initial turbulent kinetic
energy. We find the analytic explanation for these persistent
fluctuations. They are magnetic-tension-driven modes associated 
with the interaction of the sheet with the external magnetic field.
The phase speed of such modes is quite large, allowing residual
motions to persist without dissipation in the flux-freezing limit,
even as they are nonlinear with respect to the sound speed.
We speculate that long-lived large-scale MHD modes such as 
these may provide the key to understanding observed supersonic
motions in molecular clouds.

\end{abstract}


\section{Introduction}

Nonthermal linewidths are ubiquitous in molecular clouds 
\citep{sol87}  
and are interpreted to represent highly supersonic random 
internal motions \citep[see][for a recent review]{mck07}.
Principal component analysis \citep{bru09}
reveals that most of the energy is contained in modes that span 
the largest scale of the cloud.

Since molecular clouds are threaded by large-scale magnetic fields,
an attractive suggestion has been that the turbulence represents
supersonic but sub-\Alfvc magnetohydrodynamic (MHD) waves, with the
noncompressive shear \Alf mode identified as a possible long-lived component
\citep{aro75}. This was intended to bypass the usual problem of rapid
dissipation of supersonic hydrodynamic turbulence through 
shocks. A compilation of available Zeeman measurements of the
(line-of-sight) magnetic field strength, gas density $\rho$, 
and one-dimensional velocity dispersion $\sigma_v$ shows that cloud
fragments obey a direct linear correlation between $\sigma_v$ and the mean
\Alf speed $V_{\rm A}$ \citep{mye88,bas00}, with a proportionality
coefficient $\approx 0.5$. This correlation
may be attributed to a rough equality of the 
magnitudes of gravitational, magnetic, and turbulent energies, and
was interpreted by \citet{mou95} and \citet{mou06}
to mean that the motions are \Alfvc disturbances in which the perturbed
magnetic field is comparable in strength to the background magnetic 
field. Furthermore, the MHD motions in a molecular cloud
may represent long-wavelength standing waves, as argued
by \citet{mou75,mou87}. This brings out the possibility of
``global'' effects (e.g. due to cloud boundaries and external interaction) 
being important in understanding cloud turbulence, and the 
need to go beyond comparing observations with models 
of wave propagation in an infinite medium.

In direct contrast to the scenario of long-lived motions, numerical 
simulations of molecular cloud turbulence using a three-dimensional
simulation cube with periodic boundary conditions have revealed that 
supersonic MHD turbulence decays away rapidly, like its hydrodynamic 
counterpart, on about a sound crossing time of the driving scale
\citep{sto98,mac98,mac99,ost01}. This happens in either the case of
sub-\Alfvc or super-\Alfvc turbulence,
and in both cases, turbulence is maintained for long periods only 
by constant driving of velocity perturbations in Fourier space. 
When interpreting the above results, we should keep in mind that
periodic box simulations represent a ``local'' patch of uniform background 
density
that is embedded within a larger cloud, and are equivalent to 
studying an infinite uniform medium. By comparison,
a 1.5-dimensional global model including vertical stratification 
\citep{kud03,kud06} found that the decay of turbulence could be delayed, but only
mildly, by some transfer of internal kinetic energy from small to large 
scale modes along the magnetic field direction.
The rapid turbulence dissipation in all of these models
is due to the presence of shocks and takes place under the assumption of
magnetic flux freezing, without any contribution from magnetic field
dissipation, e.g., by ambipolar diffusion.

The bottom line from the above studies is that all previous numerical modeling
of MHD turbulence leads to rapid dissipation, in about a crossing time,
with a logical conclusion that matching observations requires vigorous
driving of turbulence from unspecified sources.
The alternate possibility of maintaining some long-lived
global modes is appealing but remained largely unexplored quantitatively.

In a recent paper, \citet{bas09b} carried out an extensive parameter survey of
fragmentation initiated by nonlinear turbulent flows, employing the magnetic
thin-sheet approximation and also including the effect of ambipolar diffusion
\citep[see also][]{bas04,li04,nak05,cio06,bas09a}.
In this approximation, the sheet interacts at its upper and lower
surfaces with an external magnetic field, and can be considered a global model
in the $z$-direction (parallel to the mean background magnetic field),
although it is a local (periodic) model in the $x$- and $y$-directions.
\citet{bas09b} found that initial turbulent fluctuations decayed away
quite rapidly in all models with supercritical mass-to-flux ratio, as well
as for subcritical models that included the effect of ambipolar diffusion.
However, a surprising result was that subcritical clouds evolving under
flux-freezing were able to maintain a substantial portion of their initial
kinetic energy to indefinitely large times.

In this paper, we analyze this unique instance of 
a turbulent MHD simulation that yields long-lived nonlinear motions. 
We perform a suite of numerical simulations to test its generality,
and also establish an analytic explanation for this very interesting result.

\section{Method}

The thin-sheet equations are obtained by vertically integrating the 
full equations for a partially ionized, magnetic, self-gravitating, 
isothermal fluid in the vertical direction from $z=-Z(x,y)$ to $z=+Z(x,y)$. 
Details of this integration are found in \citet{cio93} and \citet{cio06}.
The nonaxisymmetric thin-sheet equations, formulation of our model, and our
numerical methods are described in several papers \citep{cio06,bas09a,bas09b}. 
The evolution equations for the magnetized thin
sheet include the effects of magnetic tension due to the external magnetic
field $\bl{B}(x,y,z)$. It is calculated as a potential field, 
with the vertical magnetic field in the equatorial plane, $\Beq(x,y)$, 
acting as a source for $\bl{B}(x,y,z) - \Bref\hat{z}$, much as 
$\sign(x,y)$ acts as a source for the gravitational field.
Periodic boundary conditions are applied to a square Cartesian region
of size $L$.
The initial background state has a
uniform (in the $x$- and $y$- directions) neutral surface density $\signi$ and 
a uniform 
vertical ($z$-direction) magnetic field $\Bref$. We input nonlinear velocity 
fluctuations with spectrum $v_k^2 \propto k^{-4}$ in Fourier
space, where $k$ is the absolute value of the wavenumber, 
and modes are damped at a 
fixed (small) scale that is independent of the box size or the number
of grid zones used in a simulation. 

The gas is isothermal with sound speed $\cs$, and 
partial ionization is mainly due to cosmic rays.
This introduces the dimensionless free parameter $\tniitil \equiv \tnii/t_0$,
where $\tnii$ is the initial neutral-ion collision time, 
and $t_0 = \cs/2 \pi G \signi$ 
is a characteristic time in the problem. The flux-freezing limit, used 
extensively in this paper, corresponds to $\tniitil = 0$.
Another important parameter is the initial dimensionless mass-to-flux ratio
$\mui \equiv 2 \pi G^{1/2} \signi/\Bref$, i.e., $\mui > 1$ yields a supercritical
cloud in which fragmentation occurs dynamically and $\mui <1$ yields a subcritical
cloud in which fragmentation is driven on a longer time scale by ambipolar
diffusion \citep[see][]{cio06}. Turbulent initial conditions introduce the
dimensionless parameter $v_a/\cs$, where $v_a$ is the rms
amplitude of the initial velocity fluctuations in each direction.
Finally, we also vary the ratio $L/L_0$ (where 
$L_0 = \csq/2 \pi G \signi$ is a 
characteristic length scale of the system), and $N$, 
the number of grid zones in each direction.
An additional parameter, the dimensionless external
pressure $\Pexttil = 2\Pext/\pi G \signi^2$, is kept fixed at 0.1 in all models,
and does not play an important role in the dynamics.

\section{Results}

\subsection{Canonical models}
\label{canonical}

\begin{figure}
 \centering
\resizebox{\hsize}{!}{\includegraphics{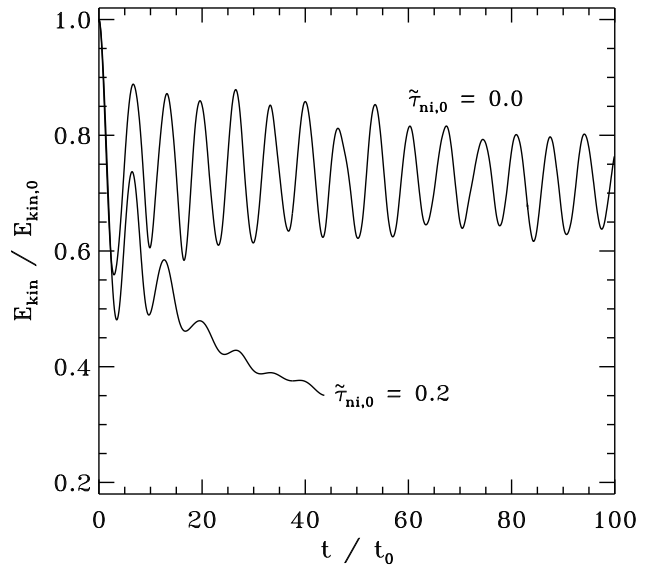}}
      \caption{Kinetic energy $\Ekin$ normalized to its initial value $\Ekini$ for
two models. Each model has $\mui=0.5$, $v_a = 2\cs$, and $L/L_0 = 32\pi$.
One model evolves with flux freezing, $\tniitil=0$, and is run with $N=512$.
The other model has partial coupling
of neutrals and ions characterized by $\tniitil=0.2$, and is run with $N=1024$.
}
         \label{fig1}
\end{figure}

Two models illustrate the key result for subcritical clouds with turbulent initial
conditions. Figure~\ref{fig1} shows the time evolution of total kinetic energy
for a model which allows for neutral-ion drift ($\tniitil=0.2$), and another 
model which has flux freezing ($\tniitil=0$).
The value $\tniitil=0.2$ corresponds to the canonical ionization fraction implied by 
primarily cosmic ray ionization:  $\chi_{\rm i} \simeq 10^{-7}(\nn/10^4\cmc)^{-1/2}$ 
\citep[see][]{tie05}, where $\nn$ is the number density of neutrals.
Both models are characterized by $\mui=0.5,v_a=2\,\cs,L=32\pi L_0$. The flux-frozen
run has $N=512$ while the ambipolar diffusion run has $N=1024$.
The evolution of the ambipolar diffusion model terminates at time $t=45.4\,t_0$, 
when the highest column density in the simulation reaches $100 \,\signi$ --- 
a useful proxy for runaway collapse of the first core.
At this time, the kinetic energy has decayed away substantially, and appears
to still be declining. In contrast, the flux-frozen model has, after an initial loss of 
some kinetic energy, stabilized to executing oscillations about a mean value
$\Ekin \approx 0.7\,\Ekini$. 
We have run simulations with flux freezing up to $t \approx 35,000\,t_0$ with no
change in this behavior.

\begin{figure}
 \centering
\resizebox{6cm}{!}{\includegraphics{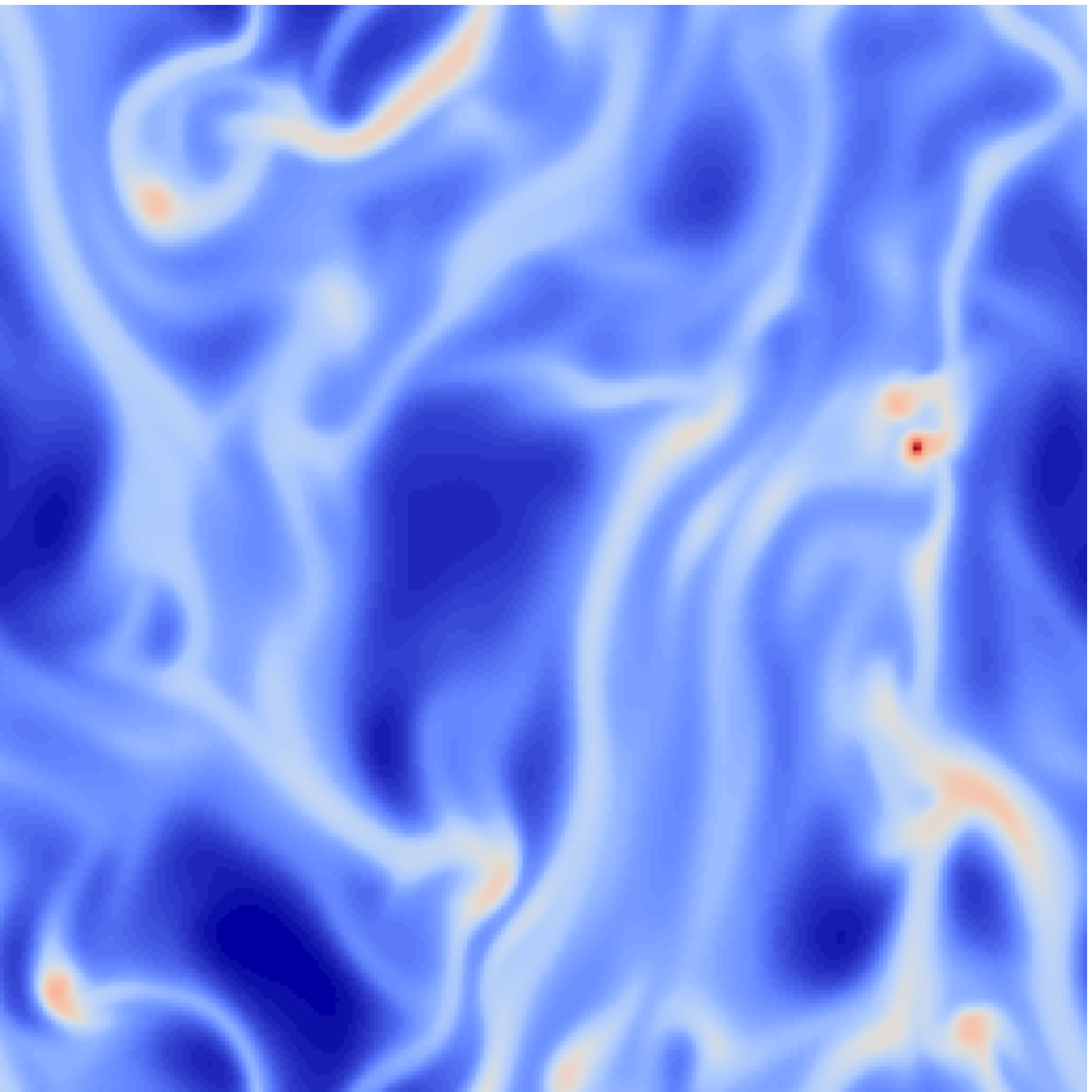}}
\resizebox{6cm}{!}{\includegraphics{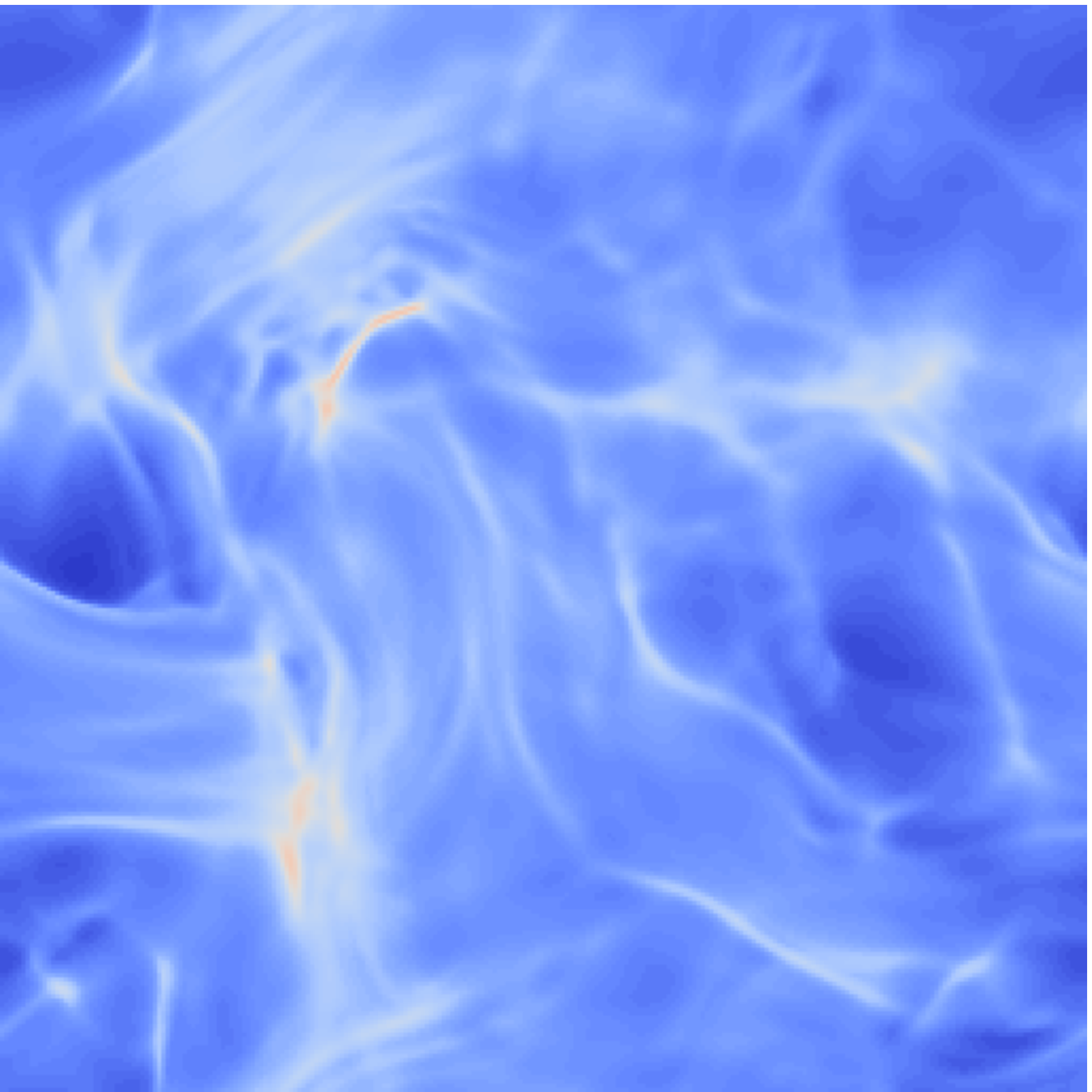}}
      \caption{Images of gas column density $\sign(x,y)/\signi$ 
for models with ambipolar diffusion (top) and flux freezing (bottom),
shown in identical color schemes that are proportional to the logarithm
of the column density. 
Both models have $\mui=0.5$, initial turbulence with
$v_a = 2\,\cs$ and spectrum $v_k^2 \propto k^{-4}$, and are run with
$N=256$. 
The ambipolar diffusion model terminates at time $t=43.5\,t_0$ in this realization 
due to the eventual runaway collapse of a core (in the upper right of the image). 
The flux-frozen model is shown at the same time, 
but continues to evolve to indefinitely large times without collapse. 
It shows a more wispy column density structure, with no evidence of 
monolithic collapse toward any density peaks. 
An animation of the evolution of each model is available online.
}
         \label{fig2}
\end{figure}

\begin{figure}
 \centering
\resizebox{6cm}{!}{\includegraphics{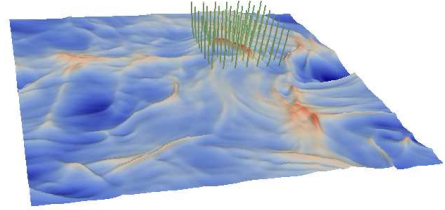}}
      \caption{Surface map of gas column density $\sign(x,y)/\signi$ 
for the same flux-frozen model that is shown in Fig.~\ref{fig2}. 
The elevation and color of the surface is proportional to the logarithm of
the local column density.
The sheet is viewed from a non-face-on viewing angle. The external
magnetic field above the sheet is also represented by field lines, 
in a region near a density peak. 
An animation of the evolution of the model, including field line
evolution, is available online.
}
         \label{fig3}
\end{figure}

Figure~\ref{fig2} shows color images of the column density for models
that are equivalent to the ones described above, but with $N=256$.
The ambipolar diffusion model is shown at its end time $t=43.5\,t_0$, and 
the flux-frozen model is shown at the 
same time, although it continues to evolve indefinitely.
The ambipolar diffusion model shows evidence of monolithic collapse toward 
one or more density peaks, while the flux-frozen model shows a more wispy character 
and gives no indication of impending collapse in any region, neither visually
nor quantitatively.
Figure~\ref{fig3} shows a model snapshot of the flux-frozen model
but viewed from a three-dimensional perspective, with the external field lines
illustrated in the region above the sheet. Since this is a subcritical
model, the field lines are not significantly deformed. The pitch angle 
of the magnetic field relative to the vertical direction, measured 
at the sheet surface, 
maintains an average value that is a little less than $10^{\circ}$. 
For a comparison of field line curvature for models with a range of 
$\mui$, see \citet{bas09a,bas09b}.
Animations of both 
Figures \ref{fig2} and \ref{fig3}, with the latter also showing external 
field line evolution, are available online.

\subsection{Connection to Linear Analysis}

To gain insight into the long-lived mode for the flux-freezing model, we revisit the modal 
analysis of a partially ionized magnetized sheet. This has been carried out
by \citet{mor91} and \citet{cio06}. 
Starting with a static uniform background, the thin-sheet equations can be
expanded by writing any physical quantity $f(x,y,t) = f_0 +
\delta f_a\,\exp[i(k_x x + k_y y - \omega t)]$, where $f_0$ is the
unperturbed state, $\delta f_a$ is the amplitude of the perturbation,
$k_x$ and $k_y$ are the wavenumbers in the $x$- and $y$- directions,
and $\omega$ is a complex angular frequency.
For assumed small perturbations, one can retain only terms up to first
order in perturbed quantities, resulting in Equations 
32(a) - 32(d) of \citet{cio06}. 
Those equations can be combined to yield the dispersion relation
\begin{eqnarray}
\label{disprel}
& (\omega + i\,\theta) \left[\,\omega^2 - \ceffsq k^2 + 2 \pi G \signi k \,\right]  \nonumber \\
& = \omega \left[\, \vai^2 k^2 + 2 \pi G \signi \, \mui^{-2} k \,\right]  ,
\end{eqnarray}
where
\beq
\theta =  2 \pi \, \tnii \left[\, \vai^2 k^2 + 2 \pi G \signi \, \mui^{-2} k \,\right]
\eeq
contains the effect of ambipolar diffusion.
In the above equations, we have introduced the \Alf speed $\vai$ for physical clarity of the 
magnetic-pressure-driven terms proportional to $k^2$ ($\equiv k_x^2 + k_y^2)$, 
while retaining $\mui$ for physical
clarity of the magnetic-tension-driven terms proportional to $k$ 
($\equiv |\sqrt{k^2}|$). The two parameters
are actually related: $\vai^2 \equiv \Bref^2/4 \pi \rhoni = 2 \pi G \signi \mui^{-2} Z_0$,
where the sheet half-thickness $Z_0 = \signi/2 \rhoni$ and $\rhoni$ is its mass volume 
density.
The quantity $\ceff$ is an effective sound speed that takes into account the restoring
force due to an external pressure $\Pext$ \citep[see][]{cio06}.
In the flux-freezing limit, Equation (\ref{disprel}) becomes
\beq
\label{disprelff}
\omega^2= (\vai^2 + \ceffsq)\, k^2 + 2 \pi G \signi \, (\mui^{-2} -1) k.
\eeq
In the case of subcritical clouds ($\mui < 1$), the second term on the
right-hand side becomes a stabilizing term rather than a destabilizing term associated
with gravitational instability.
However, the full dispersion relation (Equation (\ref{disprel})) does contain 
destabilizing terms due to ambipolar diffusion; the effect on a subcritical cloud 
is a ``slow'' instability leading to collapse on an ambipolar diffusion time scale 
rather than a dynamical time.
Equation (42) of \citet{cio06} yields a good approximation to the ambipolar diffusion
growth time for significantly subcritical clouds.

Equation (\ref{disprelff}) shows that long-wavelength modes 
evolving under flux freezing have a phase speed 
\beq
\label{vphase}
\vph \equiv \frac{\omega}{k} =  \sqrt{(\mui^{-2} - 1) G \signi  \lambda} \: ,
\eeq
where $\lambda = 2\pi/k$. These modes are driven by the 
restoring force of the magnetic tension of inclined field lines that 
connect the sheet to the external medium.
These magnetic-tension-driven modes should not be 
confused with the traditional MHD wave modes.
Within the thin sheet, in the short-wavelength limit, magnetic pressure drives the 
magnetosound mode with phase speed $\vms = (\vai^2 + \ceffsq)^{1/2}$.

Since $\vph \propto \sqrt{\lambda}$, it achieves significant values 
(much larger than $\vms$), for $\mui=0.5$ and wavelengths equal to the box 
sizes we consider: $L = 16\pi\,L_0 - 128\pi\,L_0$.
The values are in the range $4.9\,\cs - 13.9\,\cs$ and typical values 
of input parameters would   
correspond to dimensional box sizes $\approx 2 - 16$ pc 
\citep[see][for scaling of units]{bas09a,bas09b}.
Since the restoring force is provided by the external potential field that can adjust
instantaneously as the sheet evolves, the modes found in this linear analysis cannot
be applied to arbitrarily large wavelengths. In reality, there must be time for 
readjustment of the external field. The magnetic potential $\Psi_M(x,y,z)$ above and
below the sheet decays as $\exp\,(-k|z|)$ \citep[see][]{cio06,bas09a}, 
so that a characteristic height of deformation of the field lines is $k^{-1}$. 
The ratio $\epsilon$ of the \Alf crossing time across this distance 
divided by the wave period must be well below unity in order for
the potential field approximation to be valid. While the \Alf crossing time
grows more rapidly ($\propto \lambda$) with increasing wavelength than does the
wave period ($\propto \sqrt{\lambda}$), we find that $\epsilon \leq 0.26$ 
for modes of even our largest box size, if
the external density $\rho_{\rm ext} \leq 0.1\, \rhoni$.
The nature of a low-density medium external to clouds or clumps is discussed in 
Section~\ref{disc}.

An interesting analogy can be made between the magnetic-tension-driven modes and
gravity-driven waves in deep water. There, the undulations of wavenumber $k$ on a water surface can 
be felt down to a characteristic depth $k^{-1}$. Velocities below the surface are determined from
a velocity potential solution of Laplace's equation.
This is partly due to the incompressible fluid approximation in which 
the water pressure can adjust instantaneously.
A clearer mathematical analogy also occurs in the following manner.
Since the vertical gravitational field above and below a uniform 
thin sheet has magnitude
$|g_z| = 2 \pi G \signi$, Equation (\ref{vphase}) may be rewritten as
\beq
\vph = \sqrt{(\mui^{-2} -1) |g_z|/ k} \: ,
\eeq
in analogy to the phase speed $v = \sqrt{g/k}$ for deep water waves 
in a constant gravitational field $g$.

\subsection{Further modeling}

The implication of the high values of $\vph$ for the 
magnetic-tension-driven modes in the plane of a thin sheet is that
waves with nonlinear particle motions (in comparison to the sound speed $\cs$
or magnetosound speed $\vms$) may still act as linear waves since their
material motions are much slower than $\vph$. They will then evolve (in the flux-freezing limit)
without any nonlinear distortion and dissipation. All models do lose significant 
kinetic energy in an early phase, due to shocks and compression that leads to 
significant losses in an isothermal gas. Small-scale modes are prone to such decay;
for them the relevant signal speed is $\vms$, which equals $2.9\,\cs$ 
for the models presented in Section~\ref{canonical}.
However, the $v_k^2 \propto k^{-4}$ spectrum means that most of the energy
is in the largest scale mode, which can survive indefinitely, as long as the
velocity amplitude is significantly less than $\vph$, either initially or after 
some nonlinear decay. 
In the magnetic-tension-driven mode, energy is stored and released by the 
magnetic field, without losses
due to ambipolar diffusion (in the $\tniitil=0$ models) or other forms of magnetic field
dissipation. Furthermore, 
the isothermal assumption does not rob any net energy at this stage. In symmetric
oscillations, energy is lost during wave compressions and an equal amount 
gained back during wave expansions.

\begin{figure}
 \centering
\resizebox{\hsize}{!}{\includegraphics{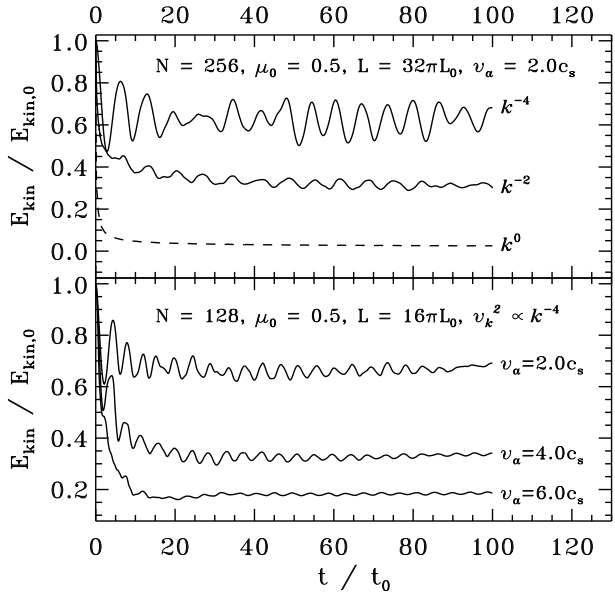}}
      \caption{Effect of different power spectra and turbulent velocity amplitudes on 
kinetic energy decay. All models evolve with flux freezing ($\tniitil=0$).
Top panel: three models with fixed $v_a$ and other parameters
(as labeled) but differing power spectra of initial turbulence, as labeled next to each
curve. The model with the most energy on the largest scale retains the greatest part of
its energy.
Bottom panel: three models with differing $v_a$ but all other parameters including
the power spectrum kept fixed, as labeled.
Models with greater $v_a$ lose a greater proportion of their initial kinetic
energy, although they retain a slightly greater absolute amount of 
kinetic energy (see Table~\ref{table1}).
}
         \label{fig4}
\end{figure}

Figure~\ref{fig4} explores the effect of different initial conditions on 
the decay and residual amount of turbulence in several flux-frozen ($\tniitil=0$) models.
The top panel shows a comparison of models with $v_k^2$ proportional to 
$k^{-4}$, $k^{-2}$, and $k^{0}$, respectively, but all having the same initial
rms speed. The spectrum with the greatest amount of energy on the largest scale 
retains the most energy, as it is the largest scale mode that has the greatest
phase speed and is most likely to survive with significant amplitude.  
The bottom panel shows that models with fixed spectrum $v_k^2 \propto k^{-4}$ but
differing $v_a$ will lose different proportions of their
initial turbulent energy. The phase speed of the largest mode in these simulation boxes 
is $\vph = 4.9\,\cs$, and increasing $v_a$ leads to greater proportionate loss of
initial kinetic energy. However, there is a weak trend toward retaining a 
greater {\it absolute} amount of energy, as a tabulation of $\vrms$,
the one-dimensional rms speed at $t \approx 100\,t_0$ in each simulation, 
reveals.
Table~\ref{table1} lists $\vrms$ 
for many models that have flux-freezing, $\mui=0.5$, 
and $v_k^2 \propto k^{-4}$ initially.
Supersonic motions remain in all models, and the residual amplitude rises
with increasing box size as well as initial velocity amplitude $v_a$.
The values of $\vrms$ appear to saturate however, so that they remain a 
reasonably small fraction (14\%-56\%) of $\vph$ for each model.

\begin{figure}
 \centering
\resizebox{\hsize}{!}{\includegraphics{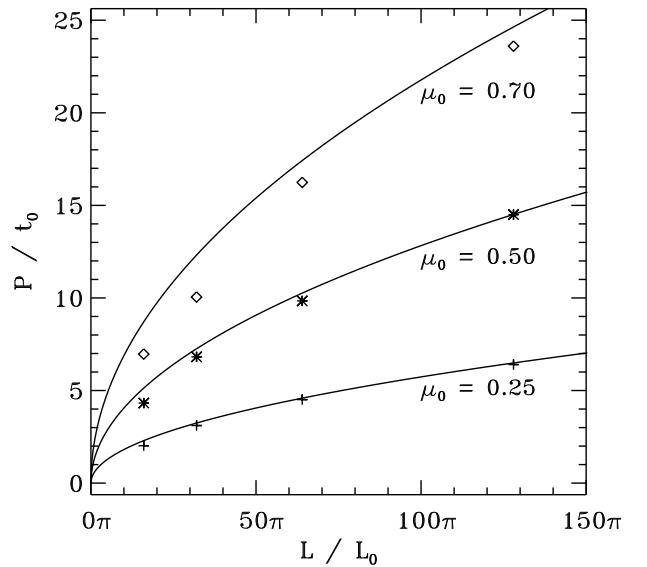}}
      \caption{Comparison of analytically predicted periods of kinetic energy oscillation
with results of simulations. Solid lines are the predicted periods from the linear theory of
large-scale flux-frozen modes driven by magnetic tension, for three different values
of the dimensionless mass-to-flux ratio $\mui = (0.25,0.5,0.7)$.  
Open diamonds represent average periods of oscillation for four models of differing
box size $L$ and fixed $\mui=0.7$.
Asterisks and plus signs represent the same but for fixed $\mui=0.5$ and $\mui=0.25$, respectively.
The agreement is between the solid lines and symbols is remarkably good, and best for 
the largest box sizes, since the analytic
prediction is made in the long-wavelength limit.
}
         \label{fig5}
\end{figure}

Oscillations of the kinetic energy are clearly visible in the
models that retain a large part of their initial energy, so that
the values of $\vrms$ in Table~\ref{table1} are varying
by up to 10\%.
The dominance of the largest mode in the initial conditions and the 
preferential nonlinear damping of smaller modes implies that the period of
the largest mode is a reasonable approximation to these observed periods
$P$. We determine $P$ by an 
average over many peak-to-peak oscillations in each model.
The kinetic energy should oscillate with half the period
of the velocity eigenfunction, so its expected period is
\beq
P =  \frac{1}{2} \frac{L}{\vph} = \frac{1}{2} \left[\frac{L}{(\mui^{-2} -1) G \signi} \right]^{1/2},
\eeq
where we have used $\lambda = L$.
In terms of the dimensionless box size $L/L_0$ and $t_0$, we can write this as
\beq
P = \frac{1}{2} \left[\frac{2\pi (L/L_0)}{\mui^{-2}-1} \right]^{1/2}\, t_0 \,.
\eeq
Figure~\ref{fig5} shows the predicted dependence in solid lines, for
$\mui = (0.25,0.5,0.7)$. Different symbols as described in 
the figure caption represent the empirical determinations 
of $P$ from various models. The agreement is remarkably good, 
and improves for the largest
box sizes, where the long-wavelength approximation made in 
Equation~(\ref{vphase}) holds particularly well.
 
\begin{table}
\begin{center}
\caption{Velocity Amplitude Results for Selected Models}
\label{table1}
\begin{tabular}{ccccccc}
\tableline\tableline
$N$ & $L/L_0$ & $\vph/\cs$ & $v_a/\cs$ & $\vrms/\cs$ & $\vrms/\vph$ \\
\tableline
128 & $16\pi$ & 4.9 & 2 & 1.7 & 0.35\\
128 & $16\pi$ & 4.9 & 4 & 2.4 & 0.48\\
128 & $16\pi$ & 4.9 & 6 & 2.6 & 0.53\\
256 & $32\pi$ & 6.9 & 2 & 1.6 & 0.23\\
256 & $32\pi$ & 6.9  & 4 & 3.4  & 0.49\\
256 & $32\pi$ & 6.9  & 6 & 3.9  & 0.56\\
512 & $32\pi$ & 6.9  & 2 & 1.7  & 0.25\\
512 & $64\pi$ & 9.8  & 2 & 1.9 & 0.19 \\
512 & $64\pi$ & 9.8  & 3 & 2.7 & 0.28 \\
512 & $64\pi$ & 9.8  & 4 & 2.9 & 0.29 \\
1024 & $128\pi$ & 13.9  & 2 & 1.9 & 0.14 \\
1024 & $128\pi$ & 13.9  & 4 & 3.4 & 0.25 \\
1024 & $128\pi$ & 13.9  & 6 & 5.0 & 0.36 \\
\tableline
\end{tabular}
\tablecomments{All models above have $\mui=0.5$, $\tniitil=0$, and
initial turbulent spectrum $v_k^2 \propto k^{-4}$. The one-dimensional
velocity dispersion $\vrms$ is measured at $t \approx 100\, t_0$ and
is present indefinitely with variability of up to 10\%.}
\end{center}
\end{table}

\section{Discussion}
\label{disc}

The study of the decay of MHD turbulence has generally been based on the modeling
of \alf, slow MHD, and fast MHD modes in media that have a uniform background.
The study of more complex global MHD (including magnetogravitational) modes for 
molecular clouds remains to be explored. 
In this paper, we have analyzed 
turbulent decay in a magnetically subcritical sheet-like cloud.
It is an idealized model of a molecular cloud that is tied to 
a magnetic field anchored in the interstellar medium.
A large fraction of the initial input kinetic energy is retained by the
deformed magnetic field, and then persists in the cloud as large-scale
oscillations. These represent linear waves of large extent which 
are nevertheless supersonic since the phase speed of the 
magnetic-tension-driven modes is up to $\approx 10$ times the 
sound speed for typical cloud sizes. 

Our model may approximate the situation of molecular clouds
that are embedded in a low-density warm H {\sc I} halo, or even the case of
molecular cloud clumps that may be embedded in a matrix of H {\sc I}
gas \citep[see][]{hen06}. Flux freezing is
a good approximation for molecular cloud envelopes (as opposed to dense cores), 
due to significant photoionization by background starlight \citep{mck89,cio95}.
Observations also reveal that the low-column-density molecular cloud
envelopes actually contain most of the cloud mass \citep{kir06,gol08}.
These envelopes may have a subcritical mass-to-flux ratio, as 
implied by their lack of star formation \citep{kir06}, velocity data 
\citep[e.g., in Taurus,][]{hey08}, and the subcritical state of the
H {\sc I} clouds \citep{hei05} from which molecular clouds are presumably assembled.

Continuous driving of turbulent motions in molecular clouds
is often invoked because the canonical numerical result of decay in a 
crossing time \citep[e.g.,][]{sto98,mac99} is inconsistent with estimated cloud lifetimes
that are at least a few crossing times \citep{wil97}. \citet{bas01}
have argued that continuous driving of turbulence is consistent
with observational constraints only if the driving occurs on the largest
scale in the cloud, i.e., most of the energy is contained on that scale.
Furthermore, continuous driving may not even be required if the decay time
of the large-scale modes is greater than or equal to the estimated cloud lifetimes.
Our models suggest that large-scale modes that are coupled to the
external magnetic field can persist for very long times, thus
reducing the need for continuous driving in order to explain observations.
These modes preferentially span the largest scales in the
model cloud, in agreement with analysis of observed cloud turbulence
\citep{bru09}. 
Future spectral line modeling of the large-scale cloud oscillations in our 
model cloud may make for useful comparison with observations, 
as has been done in 
a previous study of motions in the vicinity of dense cores
\citep{kir09}.

A counter-effect to the maintenance of large-scale modes 
is the loss of energy to the external medium.
This can be accomplished by a coupling of the magnetic-tension-driven
modes to MHD modes in the external medium.  
In a related example, \citet{eng02} found significant energy loss
to an external medium during core contraction using an approximate 
treatment of the transmission of transverse \Alf waves through 
the bounding surfaces of a thin sheet.
Some form of MHD wave coupling is certainly at work between 
any molecular cloud and its environment,
although one may also argue that a clump embedded in
a larger complex may reach a steady state in which it gains as much 
energy from its exterior as it loses. In any case, the study of the propagation
of waves outside the cloud is outside the scope of our model. 
Future three-dimensional global MHD models of molecular clouds,
which include the effect of an external medium, can address this point.

\section{Summary}
\label{summ}

We have demonstrated that MHD modes driven by the tension of inclined
magnetic field lines have a large phase speed for subcritical clouds,
which increases in proportion to the square root of the wavelength.
Numerical simulations show that nonlinear motions 
(in comparison to the sound speed) persist
indefinitely for thin-sheet evolution in the limit of
magnetic flux-freezing. These are the first of any variety of
MHD turbulence simulations that show long-lived nonlinear motions.
For the broad set of models that we study, the residual one-dimensional rms 
material motions are in the range of $\approx 2\,\cs -5\,\cs$ for 
cloud sizes in the range of $\approx 2\,\pc - 16\,\pc$.
We find that runaway collapse toward isolated density peaks occurs
when partial ionization due to (primarily) cosmic rays and ambipolar 
diffusion is included.
However, the flux-freezing results can be relevant to understanding
the low-column-density molecular cloud envelopes, which are photoionized
to the level of effective flux-freezing, and contain most of the mass
in a molecular cloud. 
For those regions, there is a particularly important role for
wave modes driven by a magnetic field that threads the cloud 
and is connected to an external medium.
Long-wavelength modes such as the ones we study may provide 
at least part of the explanation for widely observed supersonic motions 
in molecular clouds.  

\acknowledgements
We thank the referee for insightful comments.
S.B. acknowledges the hospitality of the Isaac Newton Institute for Mathematical Sciences
at Cambridge University during the writing of this paper.
S.B. was supported by a Discovery Grant from NSERC. W.B.D was supported by
an Alexander Graham Bell Canada Graduate Scholarship from NSERC.

\end{document}